\documentclass[a4paper,aps,twocolumn]{revtex4}
\begin{document}

\title{\bf{Continuity of the torsionless limit as a selection rule for gravity theories with torsion}}
\author{Luca Fabbri$^{1,2}$ and Philip D. Mannheim$^{3}$}
\affiliation{$^1$INFN \& Dipartimento di Fisica, Universit{\`a} di Bologna, Via Irnerio 46, 40126 Bologna, Italy. email: luca.fabbri@bo.infn.it \\
$^{2}$DIPTEM, Universit\`{a} di Genova, Piazzale Kennedy Pad. D, 16129 Genova, Italy.\\
$^3$Department of Physics, University of Connecticut, Storrs, CT 06269, USA.
email: philip.mannheim@uconn.edu}
\date{June 15, 2014}
\begin{abstract}
While one can in principle augment gravity theory with torsion, it is generally thought that any such torsion affects would be too small to be of consequence. Here we show that this cannot in general be the case. We show that the limit of vanishing torsion is not necessarily a continuous one, with the theory obtained in the limit not necessarily coinciding with the theory in which torsion had never been present at all. However, for a standard torsion tensor that is antisymmetric in two of its indices we have found two cases in which the vanishing torsion limit is in fact continuous, namely Einstein gravity and conformal gravity. For other gravity theories of common interest  to possess a continuous limit the  torsion tensor would need to be antisymmetric in all three of its indices.
\end{abstract}
\maketitle

\section{Introduction}

The status of torsion in gravity theory is somewhat enigmatic. While there is a rich and informative body of theoretical torsion studies in the literature (see e.g. \cite{Hehl1976}, \cite{Shapiro2002}), and while there is no known principle that would actually forbid the presence of torsion in nature, as of today there is no observational evidence that would indicate that torsion actually plays any role in the real world. Because of this it is generally thought that any torsion effects that might be present in any given theory of gravity with torsion would be too weak to be observable. In this paper we call this assumption into question by showing that the limit of vanishing torsion is not necessarily a continuous one, with the theory obtained in the zero torsion limit not necessarily coinciding with the theory in which torsion had never been present at all. We have however found two cases in which the limit is in fact continuous, namely standard Einstein gravity and a particular formulation of conformal gravity, namely that in which it is generated through radiative loop corrections in an underlying spinor theory with torsion.

To construct a metric theory of gravity one must introduce a connection $\Gamma^{\lambda}_{\phantom{\alpha}\mu\nu}$. For pure Riemannian geometry the connection is given by the Levi-Civita connection
\begin{eqnarray}
\Lambda^{\lambda}_{\phantom{\alpha}\mu\nu}=\frac{1}{2}g^{\lambda\alpha}(\partial_{\mu}g_{\nu\alpha} +\partial_{\nu}g_{\mu\alpha}-\partial_{\alpha}g_{\nu\mu}).
\label{Levi-Civita-1}
\end{eqnarray}
$\Lambda^{\lambda}_{\phantom{\alpha}\mu\nu}$ is symmetric in its $\mu$, $\nu$ indices, to thus have 40 independent components, and with it one can construct a covariant derivative operator $\nabla_{\mu}$, with the metric obeying  metricity conditions with indices sequenced here as
\begin{eqnarray}
&&\nabla_{\mu}g^{\lambda\nu}=\partial_{\mu}g^{\lambda\nu}+\Lambda^{\lambda}_{\phantom{\alpha}\alpha\mu}g^{\alpha\nu}+\Lambda^{\nu}_{\phantom{\alpha}\alpha\mu}g^{\lambda\alpha}=0,
\nonumber\\
&&\nabla_{\mu}g_{\lambda\nu}=\partial_{\mu}g_{\lambda\nu}-\Lambda^{\alpha}_{\phantom{\alpha}\lambda\mu}g_{\alpha\nu}-\Lambda^{\alpha}_{\phantom{\alpha}\nu\mu}g_{\lambda\alpha}=0.
\label{metricity-2}
\end{eqnarray}

To introduce torsion one takes the connection to no longer be symmetric on its two lower indices, and defines the Cartan torsion tensor $Q^{\lambda}_{\phantom{\alpha}\mu\nu}$ according to 
\begin{eqnarray}
Q^{\lambda}_{\phantom{\alpha}\mu\nu}=\Gamma^{\lambda}_{\phantom{\alpha}\mu\nu}-\Gamma^{\lambda}_{\phantom{\alpha}\nu\mu}.
\label{torsion-3}
\end{eqnarray}
With this antisymmetry $Q^{\lambda}_{\phantom{\alpha}\mu\nu}$ has 24 independent components. Unlike the Levi-Civita connection the torsion $Q^{\lambda}_{\phantom{\alpha}\mu\nu}$ transforms as a true rank three tensor under general coordinate transformations. In terms of the torsion tensor one defines a contorsion tensor according to 
\begin{eqnarray}
K^{\lambda}_{\phantom{\alpha}\mu\nu}=\frac{1}{2}g^{\lambda\alpha}(Q_{\mu\nu\alpha}+Q_{\nu\mu\alpha}-Q_{\alpha\nu\mu}).
\label{contorsion-4}
\end{eqnarray}
With $K^{\lambda}_{\phantom{\alpha}\mu\nu}$ one constructs the generalized connection
\begin{eqnarray}
\tilde{\Gamma}^{\lambda}_{\phantom{\alpha}\mu\nu}=\Lambda^{\lambda}_{\phantom{\alpha}\mu\nu}+K^{\lambda}_{\phantom{\alpha}\mu\nu}
\label{generalizedconnection-5},
\end{eqnarray}
to give a connection that now has 64 independent components. With this generalized connection one can construct a covariant derivative operator $\tilde{\nabla}_{\mu}$, with the metric now obeying a generalized metricity condition
\begin{eqnarray}
\tilde{\nabla}_{\mu}g^{\lambda\nu}=\partial_{\mu}g^{\lambda\nu}+\tilde{\Gamma}^{\lambda}_{\phantom{\alpha}\alpha\mu}g^{\alpha\nu}+\tilde{\Gamma}^{\nu}_{\phantom{\alpha}\alpha\mu}g^{\lambda\alpha}=0
\label{generalizedmetricity-6}
\end{eqnarray}
with respect to the connection $\tilde{\Gamma}^{\lambda}_{\phantom{\alpha}\mu\nu}$.

A torsion theory is defined to be one in which one replaces $\Lambda^{\lambda}_{\phantom{\alpha}\mu\nu}$ by $\tilde{\Gamma}^{\lambda}_{\phantom{\alpha}\mu\nu}$, with the Riemann tensor 
\begin{eqnarray}
R^{\lambda}_{\phantom{\rho}\mu\nu\kappa}
=\partial_{\kappa}\Lambda^{\lambda}_{\phantom{\alpha}\mu\nu}-\partial_{\nu}\Lambda^{\lambda}_{\phantom{\alpha}\mu\kappa}
+\Lambda^{\eta}_{\phantom{\alpha}\mu\nu}\Lambda^{\lambda}_{\phantom{\alpha}\eta\kappa}
-\Lambda^{\eta}_{\phantom{\alpha}\mu\kappa}\Lambda^{\lambda}_{\phantom{\alpha}\eta\nu}
\label{metriccurvature-7}
\end{eqnarray}
being replaced by the Riemann-Cartan tensor
\begin{eqnarray}
\tilde{R}^{\lambda}_{\phantom{\rho}\mu\nu\kappa}
=\partial_{\kappa}\tilde{\Gamma}^{\lambda}_{\phantom{\alpha}\mu\nu}-\partial_{\nu}\tilde{\Gamma}^{\lambda}_{\phantom{\alpha}\mu\kappa}
+\tilde{\Gamma}^{\eta}_{\phantom{\alpha}\mu\nu}\tilde{\Gamma}^{\lambda}_{\phantom{\alpha}\eta\kappa}
-\tilde{\Gamma}^{\eta}_{\phantom{\alpha}\mu\kappa}\tilde{\Gamma}^{\lambda}_{\phantom{\alpha}\eta\nu},
\label{curvature-8}
\end{eqnarray}
with this sequencing of indices and  use of $K_{\alpha\mu\nu}=-K_{\mu\alpha\nu}$ yielding $\tilde{R}_{\lambda\mu\nu\kappa}=-\tilde{R}_{\mu\lambda\nu\kappa}$, $\tilde{R}_{\lambda\mu\nu\kappa}=-\tilde{R}_{\lambda\mu\kappa\nu}$.  In terms of the  Levi-Civita-based  $\nabla_{\mu}$ as sequenced as per Eq. (\ref{metricity-2}), the Riemann-Cartan tensor $\tilde{R}^{\lambda}_{\phantom{\rho}\mu\nu\kappa}$ admits of the convenient decomposition
\begin{eqnarray}
\tilde{R}^{\lambda}_{\phantom{\rho}\mu\nu\kappa}&=&R^{\lambda}_{\phantom{\rho}\mu\nu\kappa}
+\nabla_{\kappa}K^{\lambda}_{\phantom{\rho}\mu\nu}-\nabla_{\nu}K^{\lambda}_{\phantom{\rho}\mu\kappa}
\nonumber\\
&+&K^{\eta}_{\phantom{\alpha}\mu\nu}K^{\lambda}_{\phantom{\alpha}\eta\kappa}
-K^{\eta}_{\phantom{\alpha}\mu\kappa}K^{\lambda}_{\phantom{\alpha}\eta\nu},
\label{decomposition-9}
\end{eqnarray}
with contractions $\tilde{R}_{\mu\kappa}=\tilde{R}^{\lambda}_{\phantom{\rho}\mu\lambda\kappa}$ and $\tilde{R}=g^{\mu\kappa}\tilde{R}_{\mu\kappa}$. The specific form given for $\tilde{R}^{\lambda}_{\phantom{\rho}\mu\nu\kappa}$ holds because the torsion tensor transforms as a tensor in a standard Riemannian space, while even as it obeys Eq. (\ref{generalizedmetricity-6}) the metric continues to obey Eq. (\ref{metricity-2}). Since the torsion tensor is independent of the metric (it cannot be expressed in terms of the metric), to construct the equations of motion in the presence of torsion one performs independent variations of the action with respect to the metric and the contorsion, according to $\delta[\tilde{R}^{\lambda}_{\phantom{\rho}\mu\nu\kappa}]
=\nabla_{\kappa}[\delta \Lambda^{\lambda}_{\phantom{\alpha}\mu\nu}]
+\delta[\nabla_{\kappa} K^{\lambda}_{\phantom{\alpha}\mu\nu}]
+\delta K^{\eta}_{\phantom{\alpha}\mu\nu}K^{\lambda}_{\phantom{\alpha}\eta\kappa}
+K^{\eta}_{\phantom{\alpha}\mu\nu}\delta K^{\lambda}_{\phantom{\alpha}\eta\kappa}
-(\kappa \leftrightarrow \nu)$, where 
$\delta \Lambda^{\lambda}_{\phantom{\alpha}\mu\nu}=(1/2)g^{\lambda\alpha}(\nabla_{\mu}[\delta g_{\nu\alpha}]+\nabla_{\nu}[\delta g_{\mu\alpha}]-\nabla_{\alpha}[\delta g_{\nu\mu}])$.  The variation yields two  tensors, an energy-momentum tensor $T^{\mu\nu}$ and a spin density tensor $\Sigma^{\lambda\mu\nu}$.  To see how things work we consider first a theory based on an arbitrary function $f(\tilde{R})$ of $\tilde{R}$ where $\tilde{R}$ is the Ricci-Cartan  scalar.

\section{Discontinuities in the equations of motion}

For the action $\tilde{I}=\int d^4x (-g)^{1/2}f(\tilde{R})$, functional variation with respect to the metric and the contorsion yields 
\begin{eqnarray}
&&\frac{1}{4}g^{\mu\nu}f(\tilde{R})
-\frac{1}{2}\left[\tilde{R}^{\mu\nu}
-g^{\mu\nu}\nabla_{\sigma}\nabla^{\sigma}
+\nabla^{\mu}\nabla^{\nu}\right]f^{\prime}(\tilde{R})
\nonumber\\
&&+\frac{1}{2}\nabla_{\lambda}\left[ (g^{\mu\nu}K^{\lambda\kappa}_{\phantom{\nu\sigma}\kappa}
+K^{\mu\lambda\nu})f^{\prime}(\tilde{R})\right]
+\frac{1}{2}K^{\mu\lambda\nu}\nabla_{\lambda}f^{\prime}(\tilde{R})
\nonumber\\
&&-\frac{1}{2}\nabla^{\mu}\left[ K^{\nu\kappa}_{\phantom{\nu\sigma}\kappa}f^{\prime}(\tilde{R})\right]
-\frac{1}{2}K^{\nu\lambda}_{\phantom{\nu\lambda}\lambda}\nabla^{\mu}f^{\prime}(\tilde{R})
\nonumber\\
&&-\frac{1}{2}\left[K^{\mu\sigma}_{\phantom{\nu\rho}\sigma}K^{\nu\rho}_{\phantom{\nu\rho}\rho}
-K^{\mu}_{\phantom{\mu}\sigma\rho}K^{\nu\rho\sigma}
+K_{\sigma\rho}^{\phantom{\sigma\rho}\mu}K^{\nu\sigma\rho}
\right]f^{\prime}(\tilde{R})
\nonumber\\
&&-\frac{1}{2}K^{\sigma\rho}_{\phantom{\nu\rho}\rho}K_{\sigma}^{\phantom{\sigma}\nu\mu}
f^{\prime}(\tilde{R})+(\mu\leftrightarrow \nu)
=\frac{1}{2}T^{\mu\nu},
\label{curvature2-10}
\end{eqnarray}
and 
\begin{eqnarray}
&&\left[K^{\gamma\alpha\beta}-K^{\gamma\beta\alpha}+g^{\beta\gamma}K^{\alpha\nu}_{\phantom{\alpha\nu}\nu}
-g^{\alpha\gamma}K^{\beta\nu}_{\phantom{\alpha\nu}\nu}\right]f^{\prime}(\tilde{R})
\nonumber\\
&&+\left[g^{\beta\gamma}\nabla^{\alpha}\tilde{R}-g^{\alpha\gamma}\nabla^{\beta}\tilde{R}\right]f^{\prime\prime}(\tilde{R})
=\Sigma^{\alpha\beta\gamma}.
\label{torsion2-11}
\end{eqnarray}
In the limit of zero torsion Eq. (\ref{curvature2-10}) reduces to  
\begin{eqnarray}
&&\frac{1}{2}g^{\mu\nu}f(R)
-\left[R^{\mu\nu}
-g^{\mu\nu}\nabla_{\sigma}\nabla^{\sigma}
+\nabla^{\mu}\nabla^{\nu}\right]f^{\prime}(R)
\nonumber\\
&&=\frac{1}{2}T^{\mu\nu},
\nonumber\\
\label{curvature2-12}
\end{eqnarray}
viz. to precisely the equation of motion that would be obtained by varying $I=\int d^4x (-g)^{1/2}f(R)$ with respect to the metric in a standard Riemannian theory.  However, if we switch the torsion off in the spin density equation we do not get zero equals zero, but instead obtain a constraint equation of the form
\begin{eqnarray}
f^{\prime\prime}(R)(g^{\beta\gamma}\partial^{\alpha}R-g^{\alpha\gamma}\partial^{\beta}R)=0.
\label{torsion2-13}
\end{eqnarray}
On contracting indices we obtain
\begin{eqnarray}
3f^{\prime\prime}(R)\partial^{\alpha}R=0.
\label{torsion2-14}
\end{eqnarray}
Thus unless $f(R)$ is such that $f^{\prime\prime}(R)$ is zero all solutions to the theory would have to obey 
\begin{eqnarray}
\partial^{\alpha}R=0,
\label{torsion2-15}
\end{eqnarray}
with the only allowed solutions to Eq. (\ref{curvature2-12}) then being ones in which the Ricci scalar is a constant. The only way to avoid this highly restrictive outcome is to have $f^{\prime\prime}(R)$ be zero, to thus allow only $f(R)=aR+b$ where $a$ and $b$ are constants, viz. to only allow a standard Einstein-Hilbert theory with a possible cosmological constant term. Hence of all the possible $f(\tilde{R})$ torsion theories that one could write down, only in the one with $f(\tilde{R})=a\tilde{R}+b$ could one continuously set the torsion to zero. Hence only for this theory could one consistently take the torsion to be weak.

To understand why we obtained this outcome, we note that in varying with respect to the torsion, the equation that we will obtain for $\Sigma^{\alpha\beta\gamma}$ will be one power lower in the torsion than the action itself is. Thus if the action contains a term linear in the torsion then the equation for $\Sigma^{\alpha\beta\gamma}$ will contain a term that will not vanish in the zero torsion limit. In general then this will give us a constraint and render the limit discontinuous.  As can be seen from  Eq. (\ref{decomposition-9}) $\tilde{R}^{\lambda}_{\phantom{\rho}\mu\nu\kappa}$ contains a term that is linear in the torsion. Thus initially we might expect that even for $f(\tilde{R})=\tilde{R}$ there should be a constraint. However all the terms in $\tilde{R}^{\lambda}_{\phantom{\rho}\mu\nu\kappa}$ that are linear in the torsion are also total derivatives. In $\int d^4x (-g)^{1/2}\tilde{R} $ they thus decouple, with the first non-trivial dependence on the torsion then being quadratic, and with no zero torsion constraint then ensuing. However for actions such as  $\int d^4x (-g)^{1/2}\tilde{R}^2$ the term that is linear in the torsion involves the product of a total derivative of the torsion and an appropriate contraction of the torsionless $R^{\lambda}_{\phantom{\rho}\mu\nu\kappa}$. This cross term is not a total derivative and thus it does not decouple from the action, and the zero torsion limit then is discontinuous. Similar considerations affect actions based on any higher power of $\tilde{R}$, and thus for any $f(\tilde{R})$ other than $a\tilde{R}+b$ the zero torsion limit will be discontinuous.

These considerations do not just affect actions that are based on functions of $\tilde{R}$. They also affect general coordinate scalar actions containing general functions $f(\tilde{R}_{\mu\kappa}\tilde{R}^{\mu\kappa})$ or $f(\tilde{R}_{\lambda\mu\nu\kappa}\tilde{R}^{\lambda\mu\nu\kappa})$ of the  Ricci-Cartan and Riemann-Cartan tensors. In fact for these particular actions there is no choice for the function $f$ for which the zero torsion limit might be continuous, since coordinate invariance itself already forces these actions to contain an even number of powers of $\tilde{R}_{\mu\kappa}$ or $\tilde{R}_{\lambda\mu\nu\kappa}$, and to thus always contain terms linear in the torsion that are not total derivatives.  

However, there is one  further case that we need to examine, one that could only possibly occur for quadratic actions, since it might be possible to obtain a term linear in the torsion that would be a total divergence for some specific combination of quadratic actions of the form $\int d^4x(-g)^{1/2}[a\tilde{R}_{\lambda\mu\nu\kappa}\tilde{R}^{\lambda\mu\nu\kappa}+b\tilde{R}_{\mu\kappa}\tilde{R}^{\mu\kappa}+c\tilde{R}^2]$ for some specific values of the $a$, $b$, and $c$ coefficients. And it turns out that there actually is one, and in fact only one, choice for the coefficients for which a cancellation does in fact occur. Specifically, following integrations by parts and the use of the identity $\nabla_{\rho}R^{\rho\alpha\beta\gamma}=\nabla^{\beta}R^{\alpha\gamma}-\nabla^{\gamma}R^{\alpha\beta}$ and its contractions, the net linear term for the combination is found to be of the form
\begin{eqnarray}
&&\int d^4x(-g)^{1/2}\bigg{[}8aK_{\lambda\mu\nu}\nabla^{\lambda}R^{\nu\mu}+2bK_{\lambda\mu\kappa}\nabla^{\lambda}R^{\mu\kappa}
\nonumber\\
&&-bK^{\lambda}_{\phantom{\lambda}\mu\lambda}\nabla^{\mu}R-4cK^{\lambda}_{\phantom{\lambda}\mu\lambda}\nabla^{\mu}R\bigg{]}~+~{\rm   surface~term}.
\nonumber
\end{eqnarray}
Thus,  the only combination for which the term linear in the torsion cancels is  the one with $a=1$, $b=-4$, $c=1$, viz. the combination $\int d^4x(-g)^{1/2}[\tilde{R}_{\lambda\mu\nu\kappa}\tilde{R}^{\lambda\mu\nu\kappa}-4\tilde{R}_{\mu\kappa}\tilde{R}^{\mu\kappa}+\tilde{R}^2]$. However, quite remarkably, we recognize this specific combination to be just the one for which the term in it that is independent of the torsion altogether, viz. $\int d^4x(-g)^{1/2}[R_{\lambda\mu\nu\kappa}R^{\lambda\mu\nu\kappa}-4R_{\mu\kappa}R^{\mu\kappa}+R^2]$, just happens to be a total divergence itself (the  Gauss-Bonnet theorem), so even this combination is not of interest \cite{footnote0}. Hence within the entire class of actions based on $\tilde{R}$, $\tilde{R}_{\mu\kappa}$ and $\tilde{R}_{\lambda\mu\nu\kappa}$, only $\int d^4x(-g)^{1/2}[a\tilde{R}+b]$ leads to a consistent zero torsion limit \cite{footnote1}.

\section{Discontinuity in the Weyl-Cartan conformal case}

To discuss the implications of conformal invariance for gravity theory (see e.g. \cite{Mannheim2006,Mannheim2012}) it is convenient to first introduce the Weyl tensor in the torsionless case, viz. 
\begin{eqnarray}
&&C_{\lambda\mu\nu\kappa}= R_{\lambda\mu\nu\kappa}
-\frac{1}{2}(g_{\lambda\nu}R_{\mu\kappa}-
g_{\lambda\kappa}R_{\mu\nu}
\nonumber\\
&&-g_{\mu\nu}R_{\lambda\kappa}+
g_{\mu\kappa}R_{\lambda\nu})
+\frac{1}{6}R^{\alpha}_{\phantom{\alpha}\alpha}(
g_{\lambda\nu}g_{\mu\kappa}-
g_{\lambda\kappa}g_{\mu\nu}).~~~
\label{trivialconformalcurvature-16}
\end{eqnarray}
This tensor has the property that unlike the behavior of $R_{\lambda\mu\nu\kappa}$ itself, under a local conformal transformation of the form $g_{\mu\nu}(x)\rightarrow \Omega^2(x)g_{\mu\nu}(x)$ the Weyl tensor transforms as $C_{\lambda\mu\nu\kappa} \rightarrow \Omega^2(x)C_{\lambda\mu\nu\kappa}$ with all derivatives of $\Omega(x)$ canceling identically. In consequence, in a Riemannian geometry the action
\begin{eqnarray}
I_{\rm W}=-\alpha_{\rm g}\int d^4x (-g)^{1/2}C_{\lambda\mu\nu\kappa}C^{\lambda\mu\nu\kappa},
\label{weylaction-17}
\end{eqnarray}
with $C_{\lambda\mu\nu\kappa}C^{\lambda\mu\nu\kappa}=R_{\lambda\mu\nu\kappa}R^{\lambda\mu\nu\kappa}-2R_{\mu\kappa}R^{\mu\kappa}+(1/3)(R^{\mu}_{\phantom{\mu}\mu})^2$ and dimensionless coupling $\alpha_{\rm g}$, is locally conformal invariant. In a Riemannian geometry the quantity $(-g)^{1/2}[R_{\lambda\mu\nu\kappa}R^{\lambda\mu\nu\kappa}-4R_{\mu\kappa}R^{\mu\kappa}+(R^{\mu}_{\phantom{\mu}\mu})^2]$ is a total divergence, so that the action can be simplified to
\begin{eqnarray}
&I_{\rm W}=-2\alpha_{\rm g}\int d^4x (-g)^{1/2}[R_{\mu\kappa}R^{\mu\kappa}-\frac{1}{3}(R^{\mu}_{\phantom{\mu}\mu})^2].
\label{weylaction-18}
\end{eqnarray}

To introduce torsion in the conformal case (see e.g. \cite{Fabbri2012}) the natural procedure would be to replace the Riemann tensor by the Riemann-Cartan tensor $\tilde{R}_{\lambda\mu\nu\kappa}$ as given in Eq. (\ref{curvature-8}), with the Weyl tensor then becoming the Weyl-Cartan tensor $\tilde{C}_{\lambda\mu\nu\kappa}=\tilde{R}_{\lambda\mu\nu\kappa}-(1/2)(g_{\lambda\nu}\tilde{R}_{\mu\kappa}-g_{\lambda\kappa}\tilde{R}_{\mu\nu}-g_{\mu\nu}\tilde{R}_{\lambda\kappa}+g_{\mu\kappa}\tilde{R}_{\lambda\nu})+(1/6)\tilde{R}(g_{\lambda\nu}g_{\mu\kappa}-g_{\lambda\kappa}g_{\mu\nu})$. To be able to maintain conformal invariance in this case we need to identify a conformal transformation law for the torsion. With $\Lambda^{\lambda}_{\phantom{\sigma}\mu\nu}$ transforming as 
\begin{eqnarray}
\Lambda^{\lambda}_{\phantom{\sigma}\mu\nu}\rightarrow \Lambda^{\lambda}_{\phantom{\sigma}\mu\nu}
+\Omega^{-1}(\delta^{\lambda}_{\mu}\partial_{\nu}+\delta^{\lambda}_{\nu}\partial_{\mu}-g_{\mu\nu}\partial^{\lambda})\Omega,
\label{qtransform-19}
\end{eqnarray}
a straightforward transformation for the torsion that takes into account its antisymmetry structure is \cite{Buchbinder1985},  \cite{Shapiro2002}
\begin{eqnarray}
Q^{\lambda}_{\phantom{\sigma}\mu\nu}\rightarrow Q^{\lambda}_{\phantom{\sigma}\mu\nu}
+q\Omega^{-1}(\delta^{\lambda}_{\mu}\partial_{\nu}-\delta^{\lambda}_{\nu}\partial_{\mu})\Omega,
\label{qtransform-20}
\end{eqnarray}
where $q$ is the conformal weight of the torsion tensor. While the specific value taken by  $q$ is not known, we note that since the torsion tensor has to have the same engineering dimension as the Levi-Civita symbol, it must have engineering dimension equal to one, with $q=1$ thus being the most natural choice. 

Moreover, when $q$ is equal to one, $\tilde{\Gamma}^{\lambda}_{\mu\nu}$ transforms as $\tilde{\Gamma}^{\lambda}_{\mu\nu}\rightarrow \tilde{\Gamma}^{\lambda}_{\mu\nu}+\Omega^{-1}g^{\lambda}_{\mu}\partial_{\nu}\Omega$, with the Riemann-Cartan tensor as given in Eq. (\ref{curvature-8}) then being conformal invariant all on its own \cite{Fabbri2012}. Consequently, for this value of $q$, and in fact for this value alone  (and not even for $q=0$),  the Cartan torsion extensions of the actions given in Eqs. (\ref{weylaction-17}) and (\ref{weylaction-18}), viz. $\int d^4x (-g)^{1/2}\tilde{C}_{\lambda\mu\nu\kappa}\tilde{C}^{\lambda\mu\nu\kappa}=\int d^4x(-g)^{1/2}[\tilde{R}_{\lambda\mu\nu\kappa}\tilde{R}^{\lambda\mu\nu\kappa}-2\tilde{R}_{\mu\kappa}\tilde{R}^{\mu\kappa}+(1/3)\tilde{R}^2]$ and $\int d^4x (-g)^{1/2}[\tilde{R}_{\mu\kappa}\tilde{R}^{\mu\kappa}-(1/3)\tilde{R}^2]$ then are locally conformal invariant too, with the conformal invariance of  $\int d^4x (-g)^{1/2}[\tilde{R}_{\mu\kappa}\tilde{R}^{\mu\kappa}-(1/3)\tilde{R}^2]$ being established directly without the need to utilize any properties of $\tilde{R}_{\lambda\mu\nu\kappa}\tilde{R}^{\lambda\mu\nu\kappa}-4\tilde{R}_{\mu\kappa}\tilde{R}^{\mu\kappa}+\tilde{R}^2$. 

Now previously we had shown that there was no combination of quadratic actions for which the zero torsion limit would be continuous. Since both of the generalized conformal actions $\int d^4x (-g)^{1/2}\tilde{C}_{\lambda\mu\nu\kappa}\tilde{C}^{\lambda\mu\nu\kappa}$ and 
$\int d^4x (-g)^{1/2}[\tilde{R}_{\mu\kappa}\tilde{R}^{\mu\kappa}-(1/3)\tilde{R}^2]$ fall into this class, and since conformal invariance expressly forces us to quadratic actions \cite{footnote2}, there is no generalized conformal action  for which the zero torsion limit would be be continuous. As we thus see, if we implement conformal invariance in the torsion case by generalizing the Weyl tensor to the Weyl-Cartan tensor, we are unable to construct a conformal invariant theory in which the zero torsion limit would be continuous. To find an alternate way to implement conformal invariance in the torsion case, one that will prove to be continuous in the limit, we turn to an approach based on spinors. And while we will need to treat the spinors themselves quantum-mechanically in the following, as far as the metric and torsion are concerned they will be treated as classical  fields, just as we have been treating them in the above \cite{footnote3}.

\section{Continuity in a spinor-based conformal case}

In order to discuss spinors in the torsion case we need to develop a vierbein formalism. To this end, instead of developing Riemannian geometry via general coordinate invariance, i.e. via invariance under local translations, one considers  invariance under local Lorentz transformations. Without any reference as yet to spinors one introduces a set of vierbeins $V^{a}_{\mu}$ where the coordinate $a$ refers to a fixed, special relativistic reference coordinate system with metric $\eta_{ab}$, with the Riemannian metric then being writable as $g_{\mu\nu}=\eta_{ab}V^{a}_{\mu}V^{b}_{\nu}$. Because the vierbein carries a fixed basis index its covariant derivatives are not given via the Levi-Civita connection alone. Rather, one needs to introduce an independent second connection known as the spin connection $\omega_{\mu}^{ab}$, with it being the derivative
\begin{eqnarray}
D_{\mu}V^{a\lambda}=\partial_{\mu}V^{a\lambda}+\Lambda^{\lambda}_{\phantom{\alpha}\nu\mu}V^{a \nu}+\omega_{\mu}^{ab}V^{\lambda}_{b}
\label{vierbeinderivative-21}
\end{eqnarray}
that transforms as a tensor under both local translations and local Lorentz transformations. If we now require metricity in the form $D_{\mu}V^{a\lambda}=0$, we find that $\omega_{\mu}^{ab}$ is no longer independent but is instead given by the antisymmetric,  24-component $-\omega_{\mu}^{ab}=V^b_{\nu}\partial_{\mu}V^{a\nu}+V^b_{\lambda}\Lambda^{\lambda}_{\phantom{\lambda}\nu\mu}V^{a\nu}$, i.e. by 
\begin{eqnarray}
-\omega_{\mu}^{ab}
&=&\frac{1}{2}(V^b_{\nu}\partial_{\mu}V^{a\nu}-V^a_{\nu}\partial_{\mu}V^{b\nu})
\nonumber\\
&+&\frac{1}{2}V^{b\alpha }V^{a\nu }(\partial_{\nu}g_{\alpha
\mu}-\partial_{\alpha}g_{\mu\nu})=\omega_{\mu}^{ba}.
\label{compatibility-22}
\end{eqnarray}

To now introduce spinors, one starts with the free massless Dirac action in flat space, viz. the Poincare invariant $(1/2)\int d^4x \bar{\psi}\gamma^{a}i\partial_{a}\psi+H. c.$, where $\gamma_a\gamma_b+\gamma_b\gamma_a=2\eta_{ab}$. To make this action invariant under local translations one introduces a $(-g)^{1/2}$ factor in the measure and replaces $\gamma^{a}\partial_{a}$ by $\gamma^{a}V^{\mu}_a\partial_{\mu}$. While the resulting action is then invariant under spacetime independent Lorentz transformations of the form $\psi\rightarrow \exp(w^{ab}\Sigma_{ab})\psi$ where  $\Sigma_{ab}=(1/8)(\gamma_a\gamma_b-\gamma_b\gamma_a)$, when the function $w^{ab}$ is taken to be spacetime dependent, to continue to maintain invariance one has to augment the action with the spin connection of Eq. (\ref{compatibility-22}), to then obtain the curved space Dirac action
\begin{eqnarray}
I_{\rm D}=\frac{1}{2}\int d^4x(-g)^{1/2}i\bar{\psi}\gamma^{a}V^{\mu}_a(\partial_{\mu}+\Sigma_{bc}\omega^{bc}_{\mu})\psi +H. c.
\label{dirac-23}
\end{eqnarray}

While this action is now both locally translation and locally Lorentz invariant, for our purposes here we note that under $g_{\mu\nu}(x)\rightarrow\Omega^2(x)g_{\mu\nu}(x)$, $V^a_{\mu}(x)\rightarrow \Omega(x) V^a_{\mu}(x)$, $\psi(x)\rightarrow \Omega^{-3/2}(x)\psi(x)$, $I_{\rm D}$ is locally conformal invariant as well. We thus get local conformal invariance for free. The reason for this is that the full symmetry of the light cone where massless particles propagate is not just the $SO(3,1)$ Lorentz group but the conformal group $SO(4,2)$ with covering group $SU(2,2)$. Since the 4-component Dirac fermion transforms as the fundamental spinor representation of the conformal group, gauging Lorentz invariance then leads to local conformal invariance as well.

To introduce torsion at the vierbein level we  replace $\omega_{\mu}^{ab}$ by a 24-component (but 48 degree of freedom) torsion-dependent spin connection $\tilde{\omega}_{\mu}^{ab}$ that obeys
\begin{eqnarray}
\tilde{D}_{\mu}V^{a\lambda}&=&\partial_{\mu}V^{a\lambda}+(\Lambda^{\lambda}_{\phantom{\alpha}\nu\mu}+K^{\lambda}_{\phantom{\alpha}\nu\mu})V^{a \nu}+\tilde{\omega}_{\mu}^{ab}V^{\lambda}_{b}=0,
\nonumber\\
-\tilde{\omega}_{\mu}^{ab}&=&-\omega_{\mu}^{ab}+V^{b}_{\lambda}K^{\lambda}_{\phantom{\alpha}\nu\mu}V^{\nu a}=\tilde{\omega}_{\mu}^{ba},
\label{torsion compatibility-24}
\end{eqnarray}
and note that $\tilde{\omega}_{\mu}^{ab}$ is  left invariant under the conformal transformations of Eqs. (\ref{qtransform-19}) and (\ref{qtransform-20}) if $q=1$. With $\tilde{\omega}_{\mu}^{ab}$ we obtain a torsion-dependent Dirac action of the form
\begin{eqnarray}
\tilde{I}_{\rm D}=\frac{1}{2}\int d^4x(-g)^{1/2}i\bar{\psi}\gamma^{a}V^{\mu}_a(\partial_{\mu}+\Sigma_{bc}\tilde{\omega}^{bc}_{\mu})\psi+H. c.
\label{torsiondirac-25}
\end{eqnarray}
Integration by parts and use of  $\gamma{^a}[\gamma^{b},\gamma^c]+[\gamma^{b},\gamma^c]\gamma{^a}=4i\epsilon^{abcd}\gamma_{d}\gamma^{5} $, $\gamma^5=i\gamma^0\gamma^1\gamma^2\gamma^3$,  and $\epsilon^{abcd}V^{\mu}_aV^{\nu}_bV^{\sigma}_cV^{\tau}_d=(-g)^{-1/2}\epsilon^{\mu\nu\sigma\tau}$ yields \cite{Shapiro2002}
\begin{eqnarray}
\tilde{I}_{\rm D}=\int d^4x(-g)^{1/2}i\bar{\psi}\gamma^{a}V^{\mu}_a(\partial_{\mu}+\Sigma_{bc}\omega^{bc}_{\mu}-i\gamma^5S_{\mu})\psi,~
\label{torsiondirac-26}
\end{eqnarray}
where $S^{\mu}=(1/8)(-g)^{-1/2}\epsilon^{\mu\alpha\beta\gamma}Q_{\alpha\beta\gamma}$. In this action we note that even if the torsion is only antisymmetric on two of its indices, the only components of it that appear in $\tilde{I}_{\rm D}$ are the four that constitute that part of it that is antisymmetric on all three of its indices. For our purposes here we note that regardless of what the value of $q$ might actually be, under the conformal transformations given in Eq. (\ref{qtransform-20})  $S^{\mu}$ is left invariant. Since the torsion-independent $I_{\rm D}$ is locally conformal invariant on its own, for any $q$ it follows that the coupling of a massless Dirac fermion to torsion as given in $\tilde{I}_{\rm D}$ is fully locally conformal invariant as well.

Now in a study of dynamics based on a fermion conformally coupled to a (torsionless) Riemannian geometry and electromagnetism with a matter action of the form
\begin{eqnarray}
I_{\rm M}=\int d^4x(-g)^{1/2}i\bar{\psi}\gamma^{a}V^{\mu}_a(\partial_{\mu}+\Sigma_{bc}\omega^{bc}_{\mu}-iA^{\mu})\psi,
\label{emdirac-27}
\end{eqnarray}
it was noted \cite{tHooft2010}, \cite{Mannheim2012} that a path integration $\int D\bar{\psi}D\psi\exp(iI_{\rm M})$ over the fermions (equivalent to a one fermion loop Feynman graph) generated an effective action of the form 
\begin{eqnarray}
I&=&\int d^4x (-g)^{1/2}C\bigg{[}\frac{1}{20}\left[R_{\mu\nu}R^{\mu\nu}-\frac{1}{3}(R^{\alpha}_{\phantom{\alpha}\alpha})^2\right]
\nonumber\\
&+&\frac{1}{3}(\partial_{\mu}A_{\nu}-\partial_{\nu}A_{\mu})(\partial^{\mu}A^{\nu}-\partial^{\nu}A^{\mu})\bigg{]},
\label{emdirac-28}
\end{eqnarray}
where the  log divergent constant $C$ is regularized  as $C=1/8\pi^2(4-D)$ in dimension $D$. Noting the similarity to $\tilde{I}_{\rm D}$, the path integration $\int D\bar{\psi}D\psi\exp(i\tilde{I}_{\rm D})$ then yields a very specific effective action \cite{Shapiro2002}
\begin{eqnarray}
I_{\rm EFF}&=&\int d^4x(-g)^{1/2}C\bigg{[}\frac{1}{20}\left[R_{\mu\nu}R^{\mu\nu}-\frac{1}{3}(R^{\alpha}_{\phantom{\alpha}\alpha})^2\right]
\nonumber\\
&+&\frac{1}{3}(\partial_{\mu}S_{\nu}-\partial_{\nu}S_{\mu})(\partial^{\mu}S^{\nu}-\partial^{\nu}S^{\mu})\bigg{]}.
\label{effectiveaction-29}
\end{eqnarray}
Since $\tilde{I}_{\rm D}$ is conformal invariant $I_{\rm EFF}$ must be too, just as can be seen. Now we note that $I_{\rm EFF}$ contains no terms that are linear in the torsion. Thus even though the fermionic action $\tilde{I}_{\rm D}$ does contain a term that is linear in the torsion, the path integration over the fermionic fields converts it into a term that is quadratic in the torsion. Finally then, since $I_{\rm EFF}$ does not contain any term that is linear in the torsion, in the zero torsion limit a gravity theory based on this $I_{\rm EFF}$  action would be continuous. Thus just like the standard Einstein-Hilbert action, for the conformal $I_{\rm EFF}$  one can consistently take the weak torsion limit.

\section{Completely antisymmetric torsion}

While not conventional and perhaps even a little contrived \cite{footnote4}, we note that if we were to take the torsion, and thus the contorsion also, to be antisymmetric on all three of their indices, then all terms linear in the torsion would cancel identically in all three of $\int d^4x(-g)^{1/2}\tilde{R}_{\lambda\mu\nu\kappa}\tilde{R}^{\lambda\mu\nu\kappa}$, $\int d^4x(-g)^{1/2}\tilde{R}_{\mu\kappa}\tilde{R}^{\mu\kappa}$, and $\int d^4x(-g)^{1/2}\tilde{R}^2$. Then for any quadratic action, and thus also for one based on the Weyl-Cartan tensor, the zero torsion limit could consistently be taken. Moreover, the same analysis extends to even higher derivative theories, since if there is no term linear in the torsion in quadratic actions, there will be none in quartic actions, and so on  Thus for any $f(\tilde{R})$, $f(\tilde{R}_{\mu\kappa}\tilde{R}^{\mu\kappa})$ or $f(\tilde{R}_{\lambda\mu\nu\kappa}\tilde{R}^{\lambda\mu\nu\kappa})$ theory, once the torsion is completely antisymmetric, the zero torsion limit could then consistently be taken.

For a torsion that is only antisymmetric on two of  its indices however, we have found two cases in which the limit of zero torsion is continuous and constraint free, namely Einstein gravity and conformal gravity. Interestingly, both of these theories are currently being used to fit gravitational data (for conformal gravity fits without any need for dark matter see \cite{Mannheim2006,Mannheim2012,Mannheim2011}), with conformal gravity even being a consistent, renormalizable, and ghost-free \cite{Bender2008,Mannheim2012} quantum theory at the microscopic level, a domain where any torsion effects might first perhaps appear.


\begin{thebibliography}{99}
\bibitem{footnote2}
Since $C_{\lambda\mu\nu\kappa}$,   $C^{\lambda\mu\nu\kappa}$ and $(-g)^{1/2}$ respectively transform as 
$C_{\lambda\mu\nu\kappa}\rightarrow \Omega^2(x)C_{\lambda\mu\nu\kappa}$,   $C^{\lambda\mu\nu\kappa}\rightarrow \Omega^{-6}(x)C^{\lambda\mu\nu\kappa}$ and $(-g)^{1/2}\rightarrow \Omega^4(x)(-g)^{1/2}$ under $g_{\mu\nu}(x)\rightarrow \Omega^2(x)g_{\mu\nu}$, only actions based on quadratic functions of the Weyl tensor  could be locally conformal invariant.
\bibitem{footnote4}
Since the torsion tensor has 24 components, only four of them are accounted for by its completely antisymmetric piece $(-g)^{-1/2}\epsilon^{\mu\alpha\beta\gamma}Q_{\alpha\beta\gamma}$, and there is no immediate reason that the other 20 components should be any less relevant than these particular four. What then makes our analysis of the Dirac equation so interesting is that even if the torsion does have its full complement of 24 components, only the four of them that are completely antisymmetric couple to the fermion.
\end{thebibliography}
\end{document}